\documentclass[twocolumn, showpacs, preprintnumbers, amsmath, amssymb]{revtex4}

\usepackage{graphicx}
\usepackage{dcolumn}
\usepackage{bm}

\begin{document}

\preprint{...}

\title{Topological Theory of Classical and Quantum Phase Transition}
\author{Tieyan Si}

\affiliation{Institute of Theoretical Physics, Chinese Academy of
Sciences, Beijing, 100080, China \\ Institute of Theoretical
Physics, Lanzhou University, Lanzhou, 730000, China }

\begin{abstract}

We presented the topological current of Ehrenfest definition of
phase transition. It is shown that different topology of the
configuration space corresponds to different phase transition, it
is marked by the Euler number of the interaction potential. The
two phases separated by the coexistence curve is assigned with
different winding numbers of opposite sign. We also found an
universal equation of coexistence curve, from which one can arrive
the phase diagram of any order classical and quantum phase
transition. The topological quantum phase transition theory is
established, and is applied to the Bose-Hubbard model, the phase
diagram of the first order quantum PT is in agreement with recent
progress.

\end{abstract}

 \maketitle

\subsection{Introduction}

A phase transition(PT) brings about a sudden change of the
macroscopic properties of a system while smoothly varying a
physical parameter. While in topology, this sudden change is
always marked by Euler number. For example, when the radii $R$ of
the mouth of a rubber bag smoothly shrinks to nothing, the Euler
number jumps from 1( Euler number of a rubber bag) to 2(Euler
number of a rubber sphere) at $R=0$. One can not help asking: does
PT bear a topological origin?

On the origin of PT, Caiani $et\;al$ proposed the topological
hypothesis\cite{caiani}\cite{casetti}: the PT is determined by a
change in the topology of the configuration space, and the loss of
analyticity in the thermodynamic observables is nothing but a
consequence of such topological change. This hypothesis has
received direct evident support from numerical and analytical
computation of the Euler characteristic in $\phi^4$ lattice
model\cite{pettini}, mean-field XY model\cite{cohen} and
k-trigonometric model\cite{angelani}. It is also supported by the
Bishop-Peyard model of DNA thermal denaturation\cite{grinza}.
Recently, it is proposed that the occurrence of a PT is connected
to certain properties of the potential energy $V$\cite{kastner}.
Franzosi, Pettini and Spinelli provided a rigorous theoretical
support to this topological
hypothesis\cite{roberto}\cite{franzosi}.

In this letter, we shall start from a more practical point to
study this topological hypothesis of PT. It will be shown that the
topological number of the critical points of the microscopic
interaction potential naturally appeared in the topological
current of Euler number. Further more, based on Ehrenfest
definition of PT, we found an universal equation of coexistence
curve, from which one can arrive the phase diagram of any order
PT. This equation holds both for classical and quantum PT. From
our definition of quantum PT, we derived the phase diagram of
superfluid-Mott-insulating quantum PT in the Bose-Hubbard model,
it is in perfect agrement with the well known phase diagram.

\subsection{Topological theory of classical phase transition}

For a standard Hamiltonian system, which is described by
$H=\sum_{i=1}^{N}\frac{1}{2}p^{2}_{i}+V(q_{1},...,q_{N})$, the
Helmoltz free energy reads\cite{roberto}
$F(\beta)=-(2\beta)^{-1}Log(\pi/\beta)-f(\beta)/\beta$, where
$\beta=1/T$ is the inverse of temperature. The first part of
$F(\beta)$ comes from the kinetic energy and the second term
$f(\beta)=\frac{1}{N}Log\int{d^{N}q\;exp[-\beta{V(\vec{q})}]}$
comes from the microscopic interaction potential $V(\vec{q})$.

As we know, a PT happens when we change the temperature or
pressure of the system. In a more practical approach, we introduce
two extra microscopic practical parameter
$\{\gamma_{1},\gamma_{2}\}$, which represent pressure, external
magnetic field, electric field, and so on. The interaction
potential $\{\gamma_{1},\gamma_{2}\}$ is generalized to
$V(\vec{q},\gamma_{1},\gamma_{2})$. In the frame work of Ehrenfest
definition, the free energy $F(\gamma_{1},\gamma_{2})$ is utterly
depend on the two extra parameters.

According to Ehrenfest definition\cite{ehrenfest} of a PT, when
the system transforms from phase A to phase B, the first order
phase transition requires that the free energy
$F(\gamma_{1},\gamma_{2})$ is continuous at the transition, i.e.,
$\delta{F}=F^{B}(\gamma_{1},\gamma_{2})-F^{A}(\gamma_{1},\gamma_{2})=0$,
but
\begin{equation}\label{dF-dF-=0}
\partial_{\gamma_{1}}\delta{F}\neq0,\;\;\;\partial_{\gamma_{2}}\delta{F}\neq0,
\end{equation}
where
$\partial_{\gamma_{1}}=\frac{\partial}{\partial{\gamma_{1}}},\;\;\partial_{\gamma_{2}}=\frac{\partial}{\partial{\gamma_{2}}}$.
For the second order transition, it requires
\begin{eqnarray}\label{2order}
&&\delta{F=0},\;\;\partial_{\gamma_{1}}\delta{F=0},\;\;\partial_{\gamma_{2}}\delta{F=0},\;\;\nonumber\\
&&\partial^{2}_{\gamma_{1}}\delta{F\neq0},\;\;\partial_{\gamma_{1}}\partial_{\gamma_{2}}\delta{F\neq0},\;\;\partial^{2}_{\gamma_{2}}\delta{F\neq0}.
\end{eqnarray}
The $p$th-order PT is characterized by a discontinuity in the
$p$th derivative of the free energy. Ehrenfest definition suggests
that the change of the free energy between the two phases
$\delta{F}=F^{B}(\gamma_{1},\gamma_{2})-F^{A}(\gamma_{1},\gamma_{2})$
plays an essential role in defining the order of PT. So we choose
the $\delta{F}$ as the fundamental order parameter field.

The order parameter of the $p$th-order PT, $\vec{\phi}$, can be
chosen as the $(p-1)$th derivative of $\delta{F}$, i.e.,
$\phi^{1}=\partial^{p-1}_{T}\delta{F},\;\phi^{2}=\partial^{p-1}_{P}\delta{F}$.
In order to investigate the topological properties of a PT, we map
the order parameter into a two dimensional sphere by introducing
an unit vector field $\vec{n}$ with
$n^{a}={\phi^{a}}/{||\phi||},\;\;n^{a}n^{a}=1,$ where
$||\phi||=\sqrt{\phi^{a}\phi^{a}}\;(a=1,2)$. It can be proved that
the Gaussian curvature on the two dimensional extra-space can be
expressed in terms of $\vec{n}$:
\begin{equation}\label{J}
J=\sum_{i,j,a,b}\epsilon^{ij}\epsilon_{ab}\frac{{\partial}n^{a}}{\partial{\gamma_{i}}}\frac{{\partial}n^{b}}{\partial{\gamma_{j}}}.
\end{equation}
In light of Duan's $\phi-$mapping topological current
theory\cite{DuanSLAC}, using
$\partial_{i}\frac{\phi^{a}}{\parallel\phi\parallel}=\frac{\partial_{i}\phi^{a}}{\parallel\phi\parallel}+\phi^{a}\partial_{i}\frac{1}{\parallel\phi\parallel}$
and the Green function relation in $\phi-$space :
$\partial_{a}\partial_{a}\ln||\phi|| =2\pi\delta^{2}(\vec{\phi}),
(\partial_{a}={\frac{\partial}{\partial\phi^{a}}})$, we can prove
that
\begin{equation}
J=\delta^2(\vec{\phi})D(\frac{\phi}{\gamma})=\delta^2(\vec{\phi})\{\phi^{1},\phi^{2}\},\label{jidelta}
\end{equation}
where
$D(\phi/q)=\frac{1}{2}{\epsilon}^{jk}{\epsilon_{ab}\partial}_j{\phi}^a{\partial_k\phi}^b$
is the Jacobian vector, in the extra-two dimensional space, it is
actually the Poisson bracket of $\phi^{1}$ and $\phi^{2}$,
$\{\phi^{1},\phi^{2}\}=\sum_{i}(\frac{\partial{\phi^{1}}}{\partial{\gamma_{i}}}\frac{\partial{\phi^{2}}}{\partial{\gamma_{j}}}-
\frac{\partial{\phi^{1}}}{\partial{\gamma_{i}}}\frac{\partial{\phi^{2}}}{\partial{\gamma_{j}}})$.

In Ehrenfest definition, the $(p-1)$th order PT requires that
$\phi^{a}\neq0,(a=1,2)$, the $p$th order PT requires
$\phi^{a}=0,(a=1,2)$. Eq. (\ref{jidelta}) suggests that the
topological current of Gaussian curvature:
$J=0,\;iff\;\vec{\phi}\neq 0;\;J\neq 0,\;iff\;\vec{\phi}=0.$ So
this topological current vanishes for the $(p-1)$th PT, and only
exist for the $p$th order PT. In other words, the Euler number of
the $(p-1)$th PT is zero, and it is not zero for the $p$th order
PT.

Therefore the topological information of the system focused on the
zero points of the vector field $\vec{\phi}$. In fact, zero points
of the vector field $\vec{\phi}$ just corresponds to the extremal
point of the variation of free energy, since the vector field is
defined by the gradient field of ${\delta}F$.

For a classical system, when the microscopic interaction potential
$V(\vec{q})$ depends on the two microscopic parameters
$\{\gamma_{1},\;\gamma_{2}\}$, the Helmoltz free energy reads
$F(\gamma_{1},\;\gamma_{2})=-(2\beta)^{-1}Log(\pi/\beta)-f(\beta,\gamma_{1},\;\gamma_{2})/\beta$.
with
$f(\beta,\gamma_{1},\;\gamma_{2})=\frac{1}{N}Log\int{d^{N}q\;exp[-\beta{V(\vec{q},\gamma_{1},\;\gamma_{2})}]}$.
Then the gradient field of free energy is
\begin{equation}\label{gradient}
\frac{\partial{F}}{\partial{\gamma_{i}}}=\frac{1}{N}
\frac{\int{d^{N}q\;exp[-\beta{V}]}\frac{\partial{V}}{\partial{\gamma_{i}}}}
{\int{d^{N}q\;exp[-\beta{V}]}}=\frac{1}{N}\langle\frac{\partial{V}}{\partial{\gamma_{i}}}\rangle.
\end{equation}
Now we see the topological current relies on the critical points
of $V$ in configuration space. Therefore it is the extremal point
of the interaction potential that decides a PT. Eq.
(\ref{gradient}) also suggests that the higher order is decided by
powers of a few basic ingredient, such as
${\partial{V}}/{\partial{\gamma_{i}}}$,
${\partial^{2}{V}}/{\partial{\gamma_{i}}}{\partial{\gamma_{j}}}$,
${\partial^{3}{V}}/{\partial{\gamma_{i}}}{\partial{\gamma_{j}}}{\partial{\gamma_{k}}}$,
and so on.

The implicit function theory shows, under the regular
condition\cite{goursat} $D(\phi/\gamma)\neq 0$, i.e.,
$\{\phi^{1},\phi^{2}\}\neq0$, we can solve the equations $\phi=0$
and derive $n$ isolated solutions:
$\vec{z}_{k}=(\gamma_{k},\gamma_{k}),\;(k=1, 2,\ldots n)$.
According to the $\delta-$function theory\cite{Schouten}, one can
expand $\delta(\vec{\phi})$ at these solutions, the topological
current of Gaussian curvature is rewritten as
\begin{equation}\label{J=sum}
J=\sum_{k=1}^{l}
\beta_{k}\eta_{k}\delta(\gamma-\gamma^{1}_{k})\delta(\gamma-\gamma^{2}_{k}),
\end{equation}
here $\beta_{k}=|W_{k}|$ is the Hopf index, $W_{k}$ is the winding
number around $z_{k}$.
$\eta_{k}=$sign$D(\phi/q)_{z_{k}}=$sign$detM_{\delta{F}}(z_{k})=$sign$\{\phi^{1},\phi^{2}\}=\pm1$
is the Brouwer degree, in which $M_{\delta{F}}(z_{k})$ is Hessian
matrix.

The winding number $\beta_{k}\eta_{k}=W_{k}$ is the generalization
of the Morse index in {Morse} theory. Its absolute value
$\beta_{k}$ measures the strength of the phase.
$\eta_{k}=$sign$\{\phi^{1},\phi^{2}\}=\pm1$, represents different
phases. For example, in the phase diagram of ultra cold atoms in
an optical lattice(FIG. 1), in the area circulated by the curve,
$\eta_{k}=$sign$\{\phi^{1},\phi^{2}\}=+1$, it represents the
Mott-Insulator phase, while outside the curve,
$\eta_{k}=$sign$\{\phi^{1},\phi^{2}\}=-1$, it represents the
superfluid phase. On the curve, $\{\phi^{1},\phi^{2}\}=0$, that is
where a quantum PT takes place.  Loosely speaking, the PT which
happened at the critical point with higher winding number is
obviously different from the process happened at the point with
lower winding number.

For a more sophisticate case, when there are many phases in the
phase diagram, the total winding number of different phases,
$\chi=\int{J}d^{2}q=\sum_{k=1}^{l}{W_{k}}$, is the Euler number.
There would be a substantial change of topology of the
configuration space, when the phase diagram jumps to a higher
order PT or a lower order PT. Such as, the manifold can be jumps
from a sphere to a torus, and to a torus with many holes, the
Euler number would jumps from $\chi=2$ to $\chi=0$, then to
$\chi=2(1-h)$ with $h$ as the number of holes of the torus. During
these process, the phase transition jumps from first order to the
second order, and to the $p$th order.

\subsection{The universal equation of coexistence curve}

The most important thing we concerns about is the phase diagram of
a PT. If we have found the coexistence curve equation, the phase
diagram can be derived by solving this equation. Our topological
theory provide us an universal coexistence curve equation:
\begin{equation}\label{poisson}
\{\phi^{1},\phi^{2}\}=0.
\end{equation}
It is an unification of the special coexistence equations of
different order PT. As all know, in quantum mechanics, if two
operator  $\hat{A}$ and $\hat{B}$ commutate with each other, i.e.,
$[\hat{A},\hat{B}]=0$, they share the same eigenfunction. So Eq.
(\ref{poisson}) reflects the same properties of different phase
operators $\phi^{1}$ and $\phi^{2}$.

In the bifurcation theory of topological current\cite{honpre}, the
topological current will split into several branches, when
$D(\phi/q)=0$, i.e., $\{\phi^{1},\phi^{2}\}=0$. It is rather
surprising that the coexistence curve in the phase diagram
happened to be the bifurcation curve of topological current.

We first take the second order PT for example to demonstrate this.
The order parameter of the second order PT is
$\phi^{1}=\partial_{T}\delta{F}$ and
$\phi^{2}=\partial_{P}\delta{F}$, substituting them into the
Jacobian vector
\begin{eqnarray}\label{D(phi/x)}
\{\phi^{1},\phi^{2}\}=\frac{\partial\phi^{1}}{\partial{T}}\frac{\partial\phi^{2}}{\partial{P}}
-\frac{\partial\phi^{1}}{\partial{P}}\frac{\partial\phi^{2}}{\partial{T}}=0,
\end{eqnarray}
and using the relations
\begin{eqnarray}\label{dtt=cp}
\partial_{T}\partial_{T}{\delta}F&=&\frac{{C_{p}^{A}}-{C_{p}^{B}}}{T},\;\;
\partial_{P}\partial_{P}{\delta}F=V(\kappa_{T}^{A}-\kappa_{T}^{B}),\nonumber\\
\partial_{P}\partial_{T}{\delta}F&=&V(\alpha^{B}-\alpha^{A}),
\end{eqnarray}
we arrive
\begin{equation}\label{D(phi/x)=cpcp-vv}
D(\phi/q)=\frac{V}{T}({C_{p}^{B}}-{C_{p}^{A}})(\kappa_{T}^{B}-\kappa_{T}^{A})-(V\alpha^{B}-V\alpha^{A}).
\end{equation}
Recalling the Ehrenfest equations
\begin{equation}\label{ehrenfest}
\frac{dP}{dT}=\frac{\alpha^{B}-\alpha^{A}}{\kappa^{B}-\kappa^{A}},\;\;\;\;
\frac{dP}{dT}=\frac{C_{p}^{B}-C_{p}^{A}}{TV(\alpha^{B}-\alpha^{A})},
\end{equation}
it is easy to verify that the equation above is in consistent with
the bifurcation condition Eq. (\ref{D(phi/x)=cpcp-vv}). So the
bifurcation equation $D(\phi/q)=0$ is an equivalent expression of
the coexistence curve equation.

For the first order PT, we chose the vector order parameter as
$\phi=\partial^{0}\delta{F}$, here $'0'$ means no derivative of the
free energy. The generalized Jacobian vector of the first order PT
with $\phi=\partial^{0}\delta{F}$ is given by
\begin{equation}\label{D(phi/x)=TT-PP}
D(\frac{\phi}{q})=(\frac{\partial{F}^{B}}{\partial{T}}-\frac{\partial{F}^{A}}{\partial{T}})
-(\frac{\partial{F}^{B}}{\partial{P}}-\frac{\partial{F}^{A}}{\partial{P}})=0,
\end{equation}
in mind of the relation $\frac{\partial{F}}{\partial{T}}=-S$ and
$\frac{\partial{F}}{\partial{P}}=V$, and considering
$D(\frac{\phi}{q})=0$, we have
\begin{equation}\label{clapeyron}
\frac{dP}{dT}=-\frac{(S^{B}-S^{A})}{(V^{B}-V^{A})}.
\end{equation}
This is just the famous Clapeyron equation.

The coexistence equation $\{\phi^{1},\phi^{2}\}=0$ also holds for
a higher-order PT. We consider a system whose free energy is a
function of temperature $T$ and magnetic field $B$, then the
Clausius-Clapeyron equation becomes $dB/dT=-\Delta{S}/\Delta{M}$.
If the entropy and the magnetization are continuous across the
phase boundary, the transition is of higher order. For the $p$th
order PT, the vector field is chosen as the $(p-1)$th derivative
of $\delta{F}$,
$\phi^{1}=\partial^{p-1}_{T}\delta{F},\;\phi^{2}=\partial^{p-1}_{B}\delta{F}$.
Substituting $(\phi^{1}, \phi^{2})$ into Eq. (\ref{poisson}), we
arrive
\begin{eqnarray}\label{D(phi/B)}
D(\phi/q)=\frac{\partial^{p}\delta{F}}{\partial{T}^{p}}
\frac{\partial^{p}\delta{F}}{\partial{B}^{p}}-
\frac{\partial\partial^{p-1}\delta{F}}{\partial{B}\partial{T}^{p-1}}
\frac{\partial\partial^{p-1}\delta{F}}{\partial{T}\partial{B}^{p-1}}=0.
\end{eqnarray}
Considering the heat capacity
$\frac{{\partial}^{2}F}{{\partial}T^{2}}=-\frac{C_{B}}{T}$ and the
susceptibility $\frac{{\partial}^{2}F}{{\partial}B^{2}}=\chi$, the
bifurcation condition $D(\phi/q)=0$ is rewritten as
\begin{equation}\label{dB/dTp}
\left[\frac{dB}{dT}\right]^{p}=(-1)^{p}
\frac{\Delta\partial^{p-2}C/\partial{T}^{p-2}}{T_{c}\Delta\partial^{p-2}\chi/\partial{B}^{p-2}}.
\end{equation}
This is in excellent agreement with Kunmar's results\cite{kunmar1}.

So different phases coexists on the the bifurcation curve, when PT
occurs, the system bifurcate into several phases, each of them is
assigned with a winding number, but the total topological number
is conserved, it is a constant which relies on the topology of
configuration space.

In fact, the Jacobian vector also provide us a new spectacles to
the scaling law\cite{kunmar}. As all known, at the critical
temperature, the correlation length $\xi$ becomes infinite, from
which the divergence of the physical parameters
arises\cite{stanley}. The topological current (\ref{J}) is
invariant under renormalization group transformation
$\vec{q'}=R(\vec{q})$. The new topological current $J'$ expressed
by the new parameter $\vec{q'}=({q^{1}}',{q^{2}}')$ share the same
fixed point with $J$, $J'=J=\infty$. This is an open question
which need further study.

\subsection{Topological quantum phase transition theory}

So far there is still no such a general topological theory of
quantum PT, in the frame work of this paper, it is easy to
establish an unified topological quantum PT theory.

The usual thermodynamic phase transition at finite temperature is
driven by thermal fluctuations, as temperature is lowered, the
thermal fluctuations are suppressed. But the quantum fluctuations
still exist and play a vital role in driving the transition from
one phase to another. At the absolute zero of
temperature\cite{sondhi}, quantum PT can be accomplished by
changing not the temperature, but some parameter in the
Hamiltonian of the system. This parameter might be the charging
energy in Josephson-junction arrays (which controls their
superconductor-insulator transition), the magnetic field in a
quantum-Hall sample, doping in the parent compound of a high-Tc
superconductor, or disorder in a conductor near its
metal-insulator transition.

The order parameter field of a quantum PT can be chosen as the
ground state energy\cite{wen}
\begin{equation}\label{Eground}
E_{g}=\frac{i}{V}ln\;Z,\;\;\;Z=\int{D\phi}e^{i\int{d}xdt\mathcal{L}(\phi)}.
\end{equation}
It plays a similar role as free energy
\begin{equation}\label{canonical}
F=-\frac{1}{\beta}ln\;Z,\;\;\;Z=Tr\;e^{-\beta{H(\gamma_i(t))}}.
\end{equation}
In fact, they are equivalent. we considering a hermitian
Hamiltonian $H(t) = H(\gamma_i(t))$ which is time-dependent
through a set of parameters $\gamma_i(t)$, i = 1, ..., n. Suppose,
that for any fixed $t$ on the time interval $[0,T]$, the spectrum
of H is non-degenerate: $H(\gamma)|n(\gamma)\rangle= E_n(\gamma)|
n(\gamma)\rangle$, with $n(\gamma)\rangle$ as an eigenstate of
Hamiltonian.

As seen above, the gradient field of the free energy plays the
keys role in a classical PT, here we also choose the order
parameter of quantum PT as
\begin{equation}\label{Qgradient}
\frac{\partial{F}}{\partial{\gamma_{i}}}=-\frac{1}{\beta}
\frac{Tr\;e^{-\beta{H(\gamma_i(t))}}\frac{\partial{H}}{\partial{\gamma_{i}}}}
{Tr\;e^{-\beta{H(\gamma_i(t))}}}=\langle\frac{\partial{H}}{\partial{\gamma_{i}}}\rangle.
\end{equation}
In mind of Feynman-Hellman theorem:
$\langle{n}(\gamma)|\frac{\partial{H(\gamma)}}{\partial{\gamma_{i}}}|
n(\gamma)\rangle=\frac{\partial{E_{n}(\gamma)}}{\partial{\gamma_{i}}},$
one can sees that the order parameter of quantum PT is actually
the gradient field of the variation of the ground state energy
$\delta{E}_{g}$.

In analogy with Ehrenfest definition, the first order quantum PT,
in which the system transforms from phase A to phase B, requires
that the ground state energy $E_{g}(\gamma_{1},\gamma_{2})$ is
continuous at the transition,
$\delta{E_{g}}=E_{g}^{B}(\gamma_{1},\gamma_{2})-E_{g}^{A}(\gamma_{1},\gamma_{2})=0$,
and
\begin{equation}\label{QdF-dF-=0}
\partial_{\gamma_{1}}\delta{E_{g}}\neq0,\;\;\;\partial_{\gamma_{2}}\delta{E_{g}}\neq0,
\end{equation}
where
$\partial_{\gamma_{1}}=\frac{\partial}{\partial{\gamma_{1}}},\;\;\partial_{\gamma_{2}}=\frac{\partial}{\partial{\gamma_{2}}}$.
For the second order quantum PT, it requires
\begin{eqnarray}\label{Q2order}
&&\delta{E_{g}=0},\;\;\partial_{\gamma_{1}}\delta{E_{g}=0},\;\;\partial_{\gamma_{2}}\delta{E_{g}=0},\;\;\nonumber\\
&&\partial^{2}_{\gamma_{1}}\delta{E_{g}\neq0},\;\;\partial_{\gamma_{1}}\partial_{\gamma_{2}}\delta{E_{g}\neq0},\;\;\partial^{2}_{\gamma_{2}}\delta{E_{g}\neq0}.
\end{eqnarray}
The $p$th-order quantum PT is characterized by a discontinuity in
the $p$th derivative of difference of ground state energy
$\delta{E_{g}}$,
\begin{eqnarray}\label{Qporder}
&&\partial^{p-1}_{\gamma_{1}}\delta{E_{g}=0},\;\;\partial^{m}_{\gamma_{1}}\partial^{p-m-1}_{\gamma_{2}}\delta{E_{g}=0},(m=1,2,...,p-1),\nonumber\\
&&\partial^{p-1}_{\gamma_{2}}\delta{E_{g}=0}.\nonumber\\
&&\partial^{p}_{\gamma_{1}}\delta{E_{g}\neq0},\;\;\partial^{m}_{\gamma_{1}}\partial^{p-m}_{\gamma_{2}}\delta{E_{g}\neq0},(m=1,2,...,p),\nonumber\\
&&\partial^{p}_{\gamma_{2}}\delta{E_{g}\neq0}.
\end{eqnarray}

In the topological quantum PT theory, we can chose the vector
order parameter $\vec{\phi}$ as the $(p-1)$th derivative of
$\delta{E_{g}}$,
\begin{equation}\label{phi}
\phi^{1}=\partial^{p-1}_{\gamma_{1}}\delta{E_{g}},\quad\;\;\phi^{2}=\partial^{p-1}_{\gamma_{2}}\delta{E_{g}}.
\end{equation}
Then the phase diagram can be obtained from the coexistence curve
equation,
\begin{equation}\label{Qpoisson}
\{\phi^{1},\phi^{2}\}=\frac{\partial\phi^{1}}{\partial{\gamma_{1}}}\frac{\partial\phi^{2}}{\partial{\gamma_{2}}}
-\frac{\partial\phi^{1}}{\partial{\gamma_{2}}}\frac{\partial\phi^{2}}{\partial{\gamma_{1}}}=0.
\end{equation}
Here $\gamma_{i}$ is the physical parameters of a quantum system.

A gas of ultra cold atoms in an optical lattice has provided us a
very good experimental observation of superfluid-Mott-insulator
phase transition\cite{greiner}. In gapped quantum spin systems, an
explicit divergence of the entanglement length appears in the
ground states\cite{martin} at a topological phase transition.
Using Matrix Products States, it was found that the localizable
entanglement possess a discontinuity at its first derivative at a
Kosterlitz-Thouless phase transition\cite{martin2}. In the
following, we shall establish a complete theory of topological
quantum PT.

This system is described by the Bose-Hubbard model
\begin{equation}\label{bose}
H=-J\sum_{{\langle}i,j\rangle}(a_{i}^{\dag}a_{j}+a_{j}^{\dag}a_{j})
+\sum_{i}\varepsilon_{i}n_{i}+\frac{U}{2}\sum_{i}n_{i}(n_{i}-1),
\end{equation}
the last term $U$ corresponding to the on site repulsion between
atoms, while the first term $J$ describes the tunnelling of atoms
between neighboring sites.

Using mean-field approaches, the ground state energy of
Bose-Hubbard model can be generally expressed as the functional of
$\varepsilon_{i}$, $J$ and $U$, i.e.,
$E_{g}=E_{g}(\varepsilon_{i}, J,U).$ Following perturbation theory
up to the second order, the variation of ground state
energy\cite{oosten} $\delta{E_{g}}$ is
\begin{equation}\label{c2}
\delta{E_{g}}=[\frac{g}{{U}(g-1)-J}+\frac{g+1}{J-{U}g}+1],
\end{equation}
From our definition of the first order phase transition, the
boundary between the superfluid and the Mott insulator phases
should be decided by the equation $\delta{E_{g}}=0$. We have
plotted the phase diagram(FIG. 1).
\begin{figure}
\begin{center}\label{no1}
\includegraphics[width=0.26\textwidth]{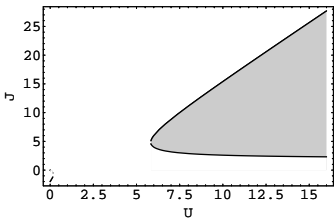}
\caption{The phase diagram of the Bose-Hubbard Hamiltonian for
g=1. Inside the curve is the Mott-insulating phase, outside the
curve is the superfluid phase.}
\end{center}
\end{figure}
This phase diagram is in perfect agrement with the well-known
phase diagram of superfluid-Mott-insulator phase
transition\cite{oosten}. It is also the same as the phase boundary
obtained from the minimizing the free energy\cite{chen}, when
treat the hopping term as perturbation. So most of the recent
phase diagram of the quantum phase transition are focused on the
first order.

As for the $p$th-order quantum PT, it is characterized by a
discontinuity in the $p$th derivative of the variation of ground
state energy. The order parameter field of the $p$th order Quantum
PT can be chosen as the (p-1)th derivative of $\delta{E}_{g}$ with
respect to $U$ and $J$, i.e.,
$\phi^{1}=\partial^{p-1}_{J}\delta{E}_{g},\;\phi^{2}=\partial^{p-1}_{U}\delta{E}_{g}$.

The equation of coexistence curve is given by
$\{\phi^{1},\phi^{2}\}=D(\frac{\phi}{\gamma})=0$,
\begin{eqnarray}\label{D(E/JU)}
\{\phi^{1},\phi^{2}\}=\frac{\partial^{p}{\delta{E}_{g}}}{\partial{J}^{p}}
\frac{\partial^{p}{\delta{E}_{g}}}{\partial{U}^{p}}-
\frac{\partial\partial^{p-1}{\delta{E}_{g}}}{\partial{U}\partial{J}^{p-1}}
\frac{\partial\partial^{p-1}{\delta{E}_{g}}}{\partial{J}\partial{U}^{p-1}}=0,
\end{eqnarray}
here we have taken $\gamma_{1}=U,\;\gamma_{2}=J$.  From this
equation, one can arrive the phase diagram of the $p$th order
quantum PT from the equation above, and find some new quantum
phases.

\subsection{conclusion}
In summary, when a PT jumps from lower order to higher order, the
Euler number of the extra configuration space would jump from one
integer to another. In other word, different order of PT
corresponds to different manifold with different Euler number.

We have found an universal equation of coexistence curve equation,
$\{\phi^{1},\phi^{2}\}=0$, from which one can derive the phase
diagram of any order. In the phase diagram, different phase are
assigned with different winding numbers, the sum of these winding
numbers is Euler number which bifurcated on the coexistence curve.

In analogy with Ehrenfest definition of classical PT, we proposed
the higher order quantum PT, in which the ground state energy
correction $\delta{E_{g}}$ plays the key role. The universal
equation of coexistence curve equation, $\{\phi^{1},\phi^{2}\}=0$,
also holds for quantum PT.

Now we proposed the general process to a obtain the phase diagram
of $p$th order quantum PT. First, we calculate the ground state
energy correction using perturbation theory. Then, substitute
$\delta{E_{g}}$ into $\{\phi^{1},\phi^{2}\}=0$, and solve this
equation. This method is an universal theory, it can be
generalized to a great variety of classical and quantum many body
system, such as t-J model, Hubbard model, and so on. It can help
us to predict some new quantum PT.

This work was supported by the National Natural Science Foundation
of China.

\end{document}